\documentclass{article}
\usepackage{graphicx} 
\usepackage{color} 
\usepackage{amsmath}
\usepackage{float}

\numberwithin{equation}{section}

\usepackage[ruled,linesnumbered]{algorithm2e}
\usepackage{appendix}
\usepackage[]{hyperref}
\usepackage[figure]{hypcap}
\usepackage[hyphenbreaks]{breakurl}
\title{Solve Mismatch Problem in Compressed Sensing} 
\author{Le Yang}

\makeatletter
\newcommand{\RemoveAlgoNumber}{\renewcommand{\fnum@algocf}{\AlCapSty{\AlCapFnt\algorithmcfname}}}
\newcommand{\RevertAlgoNumber}{\algocf@resetfnum}
\makeatother

\usepackage{cite}

\begin{document}
\maketitle
This article proposes a novel algorithm for solving mismatch problem in compressed sensing. Its core is to transform mismatch problem into matched by constructing a new measurement matrix to match measurement value under unknown measurement matrix. Therefore, we propose mismatch equation and establish two types of algorithm based on it, which are matched solution of unknown measurement matrix and calibration of unknown measurement matrix. Experiments have shown that when under low gaussian noise levels, the constructed measurement matrix can transform the mismatch problem into matched and recover original images. The code is available: \url{https://github.com/yanglebupt/mismatch-solution}.

\section{Introduction}

The basic formula of compressed sensing (CS) is 
\hyperref[eq:ori]{Eq.\ref{eq:ori}}, where $A\in R^{MXN}$ is measurement matrix, $x\in R^{NX1}$ is unknown image, $y\in R^{MX1}$ is measurement value, and $\epsilon\in R^{MX1}$ is system noise.

\begin{equation}
\label{eq:ori}
y=speckle\_measure(A\rightarrow x)=Ax + \epsilon
\end{equation}

Measurement and reconstruction are the two core steps in CS. There are a lot of traditional sparsity-regularized-based, image-CS and deep learning methods have been proposed as introduced in \cite{WOS:000497434700009, WOS:000765983700001} for reconstruction.
Traditional compressed sensing algorithms require matched pair $(y, A)$ to recover original image $x^*$, while deep learning can fit measurement matrix $A$ by training network using large number of matched pairs $(x,y)$ and then directly reconstruct original image $x^*$ from measurement value. And enhance the diversity of model fitting, matched pairs should under different measurement matrixs for mixed training\cite{WOS:000447853100004, WOS:000802187100079}. So the measurement values obtained from the unknown measurement matrix not in the training can also be directly reconstructed through the network. For measurement, most commonly used is random gaussian matrix which is signal independent and ignore the characteristics of the signal. So recently, scholars have proposed using deep learning to learn measurement matrix\cite{WOS:000497434700009, wu2019deep}.
~\\\\
Mismatch problem in compressed sensing can be seen as a type of problem. The measurement matrix and measurement value of such problems don't correspond within system noise error. But traditional compressed sensing algorithms can only solve original image according to the matched pairs, which measurement matrix and measurement value correspond within the system error range. Deep learning with mixed training and local average of measurement matrix\cite{WOS:000530854700091} can solve mismatch problem to a certain extent, which requires the measurement values during reconstruction is under the same type of measurement matrix in training, implying the correlation should not deviate too much.
~\\\\
Therefore our idea is to transform mismatch problem into matched by constructing a new measurement matrix to match measurement value under unknown measurement matrix. And then look for solution with traditional compressed sensing algorithms. We proposes several different methods to construct the new measurement matrix and provides the theoretical explanations and references for future research.

\section{Methods}
\subsection{Matched Solution of Unknown Measurement Matrix}

Measurement $y=A_u x+\epsilon_1$ under unknown measurement matrix $A_u$ . It is not possible to solve the original signal $x$ solely by relying on the measurement value $y$. So our goal is to construct a new matrix $A_{recv}$ to satisfy following optimization problem

\begin{equation}
\label{eq:goal}
A_{recv} = \arg\min_{\Lambda} ||y- \Lambda x||_2^2
\end{equation}

Then $(y, A_{recv})$ pair can be input into the compressed sensing algorithm $G$ for solution $x^{*}=G(y, A_{recv})$. This type of method is named \emph{Matched Solution of Unknown Measurement Matrix}.
~\\\\
Assuming we have another measurement $y_0=Ax + \epsilon_2$ under known measurement matrix $A$, and without nosie ($\epsilon_1=0, \epsilon_2=0$). General solution of \hyperref[eq:goal]{Problem.\ref{eq:goal}} is given by \textbf{Mismatch Equation} as \hyperref[eq:Arecv]{Eq.\ref{eq:Arecv}}. More details are provided in \hyperref[appe:A]{Appendix.\ref{appe:A}}

\begin{equation}
\label{eq:Arecv}
A_{recv}^{(y_0, y)} = \frac{1}{y_0^T\Sigma y_0}yy_0^T\Sigma A
\end{equation}

Where $\Sigma \in R^{MXM}$ can be any non zero matrix. It means relationship between measurement value and measurement matrix is not one-to-one when $x$ is fixed, indicating that for the under unknown measurement problem, there is not only one solution $A_{recv}$ transform the problem into a matched situation. So $A_{recv}^{(y_0, y)}$ is not equal to $A_u$. However, using mismatch equation alone has two drawbacks

\begin{itemize}
\item If we use the unknown measurement matrix $A_u$ measures other image $x'$, we get $y' = A_u x'$. The $(y', A_{recv}^{(y_0, y)})$ pair is still mismatched.
\item Considering noise, Mismatch Equation does not guarantee that ($y$, $A_{recv}^{(y_0, y)}$) is still matched within the error range.
\end{itemize}

Therefore, we developed an \textbf{Iterative Algorithm} as described in \hyperref[algo:1]{Algorithm.\ref{algo:1}} to get $A_{recv}$ making up for above drawbacks.


\begin{algorithm}[!h]
	\caption{$A_{recv}$ for $y'$}
	\label{algo:1}
	\LinesNumbered
	\KwIn{$y'$}
	\KwOut{$A_{recv}$}
        Parameters $A, \; PM_{image}, \; epoch$ \\
        $y_0=A * PM_{image}$ \\
        Initialize $A_{recv}=0$ \\
        $e_y \leftarrow y'$ \\ 
        $\Sigma \leftarrow (AA^T)^{-1}$ \\
	\For{$i\leftarrow 1 \; to \; epoch$}{
          $A_{recv}^{e_y} \leftarrow \frac{1}{y_0^T\Sigma y_0}\;e_y\;y_0^T\Sigma A$ \\ 
	   $A_{recv} \leftarrow A_{recv} + A_{recv}^{e_y}$ \\
          $e_y$ = $y'$ - $speckle\_measure$($A_{recv} \rightarrow x'$)
	}
\end{algorithm}

In this algorithm, $y'$ is unknown measurement value for unknown image $x'$, And we take a special solution $\Sigma=(AA^T)^{-1}$ (Derivation is provided in \hyperref[appe:A]{Appendix.\ref{appe:A}}) for mismatch equation which $A$ can be any known measurement matrix. In addition, known measurement is no longer about unknown image, but about a known image $PM_{image}$, we name this step $y_0=A * PM_{image}$ as \textbf{Pre-Measure}, which is completely known and computable.
~\\\\
In each iteration, the measurement error $e_y$ is taken into mismatch equation and accumulate it's result $A_{recv}^{e_y}$ onto $A_{recv}$. Then use updated known measurement matrix to measure image $x'$ to obtain the current measurement value. Finally update $e_{y}$ as the result of known target measurement value $y'$ minus current measurement value. Convergence proof of iterative algorithm is provided in \hyperref[appe:B]{Appendix.\ref{appe:B}}. However, this algorithm still has two disadvantages

\begin{itemize}
\item For each different unknown image, we need to run the algorithm again to obtain a new $A_{recv}$.
\item It introduces many additional measurements.
\end{itemize}

The main reason of first disadvantage is that constructed $A_{recv}$ by iterative algorithm is not equal to $A_u$, it is only a matched solution of unknown measurement $y'$. As for how to construct the real $A_{recv}$ for any unknown images, we will introduce in the next section. To reduce additional measurements, we need to use \textbf{Multiplier Property} of mismatch equation as  \hyperref[eq:m]{Eq.\ref{eq:m}}.

\begin{equation}
\label{eq:m}
\begin{aligned}
A_{recv}^{e_y}x' = \frac{1}{y_0^T\Sigma y_0}e_yy_0^T\Sigma A x' = k(x')*e_y \\
\end{aligned}
\end{equation}

\begin{equation*}
k(x')=\frac{y_0^T\Sigma A x'}{y_0^T\Sigma y_0} \in R  
\end{equation*}

This means that when $A_{recv}^{e_y}$ measure different images, the results are proportional, and the coefficient $k$ is only related to $(y_0, x')$ not related to $e_y$. Because of linear relationship of accumulation, we can use an initial $A_{recv}^{y_0}$ to measure a known image $PM_{image}$ and then measure the unknown image $x'$ calculating the scale coefficient $k(x')/k(PM_{image})$. With the scale coefficient, there is no need to measure unknown image $x'$. Therefore, we can obtain an improved iterative algorithm as described in \hyperref[algo:2]{Algorithm.\ref{algo:2}}.

\begin{algorithm}[!h]
	\caption{Matched Solution of Unknown Measurement Matrix}
	\label{algo:2}
	\LinesNumbered
	\KwIn{$y'$}
        \KwOut{$A_{recv}$}
        Parameters $A, \; epoch$ \\
	$(y_0, PM_{image}, A_{recv}) \leftarrow Construct\_Initial(A)$ \\
        $\Sigma \leftarrow (AA^T)^{-1}$ \\
        $y = speckle\_measure(A_{recv} \rightarrow x')$ \\
        $y_{pm} = A_{recv} * PM_{image}$ \\
        $k \leftarrow y / y_{pm} $ \\
        $e_y \leftarrow y' - k*y_{pm}$ \\
        \For{$i\leftarrow 1 \; to \; epoch$}{
            $A_{recv} \leftarrow A_{recv} + \frac{1}{y_0^T\Sigma y_0}\;e_y\;y_0^T\Sigma A$ \\
            $e_y$ = $y'$ - $k*A_{recv}*PM_{image}$
        }
\end{algorithm}

\begin{algorithm}[!h]
        \renewcommand{\thealgocf}{2.1}
	\caption{Construct Initial $A_{recv}^{y_0}$}
	\LinesNumbered
        \label{algo:pm}
	\KwIn{$A$}
	\KwOut{$y_0, PM_{image}, A_{recv}$}
        Parameters $PM_{image}, \; epoch$ \\
        $y_0=A * PM_{image}$ \\
        
        $e_y \leftarrow y_0$ \\ 
        $\Sigma \leftarrow (AA^T)^{-1}$ \\
        Initialize $A_{recv}=0$ \\
	\For{$i\leftarrow 1 \; to \; epoch$}{	
	   $A_{recv} \leftarrow A_{recv} + \frac{1}{y_0^T\Sigma y_0}\;e_y\;y_0^T\Sigma A$ \\
          $e_y$ = $y_0$ - $A_{recv} * PM_{image}$
	}
\end{algorithm}

We first generated an $A_{recv}=A_{recv}^{y_0}$ using $Construct\_Initial$ step in \hyperref[algo:pm]{Algorithm.\ref{algo:pm}}. Theoretically there is no need for iteration here, but using float32 to calculate the matrix will introduce precision error, so iteration is still carried out. Then we can calculate the proportion coefficient $k(x')/k(PM_{image})$ based on this matrix. But there will be noise in the measurement, so we can only obtain an approximation

\begin{gather*}
y=speckle\_measure(A_{recv}^{y_0} \rightarrow x')=k(x')y_0+\epsilon \\
y_{pm}=k(PM_{image})y_0 \\
k=\frac{k(x')y_0+\epsilon}{k(PM_{image})y_0}\approx\frac{k(x')}{k(PM_{image})} \\
\end{gather*}

After obtaining the coefficient approximation, there is no need to measure unknown image in the subsequent iteration process, and simply replace as follows

\begin{gather*}
\begin{aligned}
k*A_{recv}^{e_y}*PM_{image} &= \frac{k(x')y_0+\epsilon}{k(PM_{image})y_0} * k(PM_{image})e^y \\
&=k(x')e^y+\frac{e^y}{y_0}\epsilon \\
&=k(x')e^y+\epsilon^{'} \\
&=speckle\_measure(A_{recv}^{e_y} \rightarrow x')
\end{aligned}
\end{gather*}

This means that after the replacement, the noise is changed from $\epsilon$ to $\frac{e^y}{y_0}\epsilon$ which is decreasing, Therefore, the iterative algorithm can still converge.

\subsection{Calibration of Unknown Measurement Matrix}
\label{sec:b}
In the previous section, we provide a matched solution for the unknown measurement matrix, but it still has one disadvantage. For each different unknown image, need to run the algorithm again to obtain a new $A_{recv}$. Therefore, this section attempts to further provide a calibration method for the unknown measurement matrix. Our goal is to solve real unknown measurement matrix $A_u$ so that one calibration can be applied to any unknown image, which is an exact solution. This type of method is named \emph{Calibration of Unknown Measurement Matrix}.
~\\\\
Let's assume that unknown image can be linearly represented by orthogonal base images.
$$x'=\sum_{i=1}^{N} b_i\bf{x_i}$$

Perform Pre-Measure and Unknown-Measure on all base images
\begin{gather*}
y_{i}^0=A\mathbf{x_i}\\
y_{i}=A_u\mathbf{x_i}+\epsilon_{i}     
\end{gather*}

As proofed in \hyperref[appe:C]{Appendix.\ref{appe:C}}, when special solution $\Sigma$ in mismatch equation satisfy following condition.

\begin{equation}
\label{eq:condition}
Y\Sigma Y^T=E
\end{equation}

$E$ is the identity matrix, $Y\in R^{N \times M}$ consists of the pre-measurement value $\{y_j^0\in R^{1 \times M}\}_{j=1}^{N}$ of all base images in rows. In this case, we have a solution of $A_u$ as \hyperref[eq:es]{Eq.\ref{eq:es}} within the error range

\begin{equation}
\label{eq:es}
A_{recv}=\sum_{j}^{N} A_{recv}^{(y_{j}^0,y_j)}
\end{equation}

Let's further discuss condition \hyperref[eq:condition]{Eq.\ref{eq:condition}}. Since $N\gg M$, Y is column full rank, it is difficult to find a solution that satisfies the condition in this situation. $N$ is number of base images, we must reduce it to less than or equal to $M$ in order to have a $\Sigma$ solution. Divide the situation into the following two types

\subsubsection{Unknow Images in N-dim Space}
\label{sec:b1}
$N$ is the dimension of an unknown images. Using $N$ orthogonal base images can represent any unknown images. But as mentioned above, there is no solution for condition \hyperref[eq:condition]{Eq.\ref{eq:condition}} because of $N\gg M$.
~\\\\
We can divide all base images without overlap into $K$ groups $\{\bf{G_{k}}\}_{k=1}^{K}$, with the number of each group $\{N_{k}\}_{k=1}^{K}$ is less than or equal to $M$. Then use the base images of each group to 
calculate the corresponding $\Sigma_{k}$ that satisfy the condition. For each group $Y\in R^{N_{k}\times M}$ is row full rank, so $\Sigma_{k}$ can be easily solved by pseudo-inverse as following. $\dagger$ is the marker of pseudo-inverse.
\begin{gather}
\label{eq:consol}
\begin{aligned}
\Sigma &= Y_{right}^{\dagger}(Y^T)_{left}^{\dagger} \\
&=Y^T(YY^T)^{-1}(YY^T)^{-1}Y 
\end{aligned}
\end{gather}
And then calculate the $A_{recv}^{k}$ of each group by \hyperref[eq:es]{Eq.\ref{eq:es}}. At this point, For the unknown image $x'$, we can also perform the same grouping representation.
\begin{gather*}
x' = \sum_{k=1}^{K} x_{k} \\
x_{k} = \sum_{\bf{x_i} \in G_{k}} b_i\bf{x_i} 
\end{gather*}
If don't consider noise, we can obtain the following equation holds
\begin{gather*}
y' = A_ux' = \sum_{k=1}^{K} A_ux_{k} = \sum_{k=1}^{K} A_{recv}^{k}x_{k} \\
y_{recv} = A_{recv} x' = A_{recv} \sum_{k=1}^{K} x_{k}
\end{gather*}
$A_{recv}$ is the target matrix to be constructed. In order to make $y'=y_{recv}$, we have
\begin{gather}
\label{eq:rrrr}
A_{recv} \sum_{k=1}^{K} x_{k} = \sum_{k=1}^{K} A_{recv}^{k}x_{k} 
\end{gather}

It is also very difficult to separate $A_{recv}$ from \hyperref[eq:rrrr]{Eq.\ref{eq:rrrr}}. But we can choose special base images to make it easier. The simplest choice is $\bf{x_i} = (0,0,...,0,1,0,...0,0)$, the $i\text{-}th$ position is 1, and the rest are all 0. Base image group is $$\bf{G_{k}=\{x_i\}_{i=kM-M}^{min(kM,N)-1}}$$
In this case, $A_{recv}^{k}x_{k}$ just use $range_{k}=[kM-M, min(kM,N)-1]$ columns because of $x_{k}$ within this range is not equal zero, out of this range is zero. So we can get final $A_{recv}$ as traverse $k$ from 1 to $K$, do $A_{recv}(range_k) = A_{recv}^{k}(range_k)$.

\subsubsection{Unknow Images in M-dim Space}
\label{sec:b2}
Assuming the target unknown $N$-dimensional images can be represented by $M$ base images. The dimension of $Y$ obtained from pre-measure of the base image is $R^{M \times M}$. We can easily get a solution for condition as \hyperref[eq:condsol]{Eq.\ref{eq:consol}}.
$$x = Qb$$
$x\in R^{N\times 1}$ is unknown images, $Q\in R^{N\times M}$ is consisted of $M$ base images by columns, which is row full rank. $b\in R^{M\times 1}$ is the coordinate. If target unknown images $x$ in the space composed of column vectors of $Q$, coordinate $b$ has a unique solution. If not in, coordinate $b$ has only one least squares solution, which will further introducing additional error in \hyperref[eq:y-error]{Eq.\ref{eq:y-error}}. Now we can summarize calibration of unknown measurement matrix when unknown images is in $M$-space as described in \hyperref[algo:calidate-mspace]{Algorithm.\ref{algo:calidate-mspace}}.
\RevertAlgoNumber
\begin{algorithm}[!h]
        \renewcommand{\thealgocf}{3}
	\caption{Calibration of Unknown Measurement Matrix}
	\label{algo:calidate-mspace}
	\LinesNumbered
	\KwOut{$A_{recv}$}
        Parameters $A$ \\
        $Q \leftarrow qr\_decomposition(A)$ \\ 
        $Y \leftarrow AQ$ \\
        $Y_u \leftarrow speckle\_measure(A_u, Q)$ \\
        $\Sigma \leftarrow (Y^T)^{\dagger}Y^{\dagger}$ \\
        Initialize $A_{recv}=0$ \\
	\For{$i\leftarrow 1 \; to \; M$}{	
          $y_0 = Y(:,i)$\\
          $y=Y_u(:,i)$\\
	   $A_{recv} += \frac{1}{y_0^T\Sigma y_0}\;y\;y_0^T\Sigma A$ \\
	}
\end{algorithm}
\newpage

\section{Experiments}
\subsection{Exps in the Device with Medium Precision}
\begin{figure}[htbp]
  \centering
  \setlength{\abovecaptionskip}{-0.05cm}
  \includegraphics[scale = 0.35]{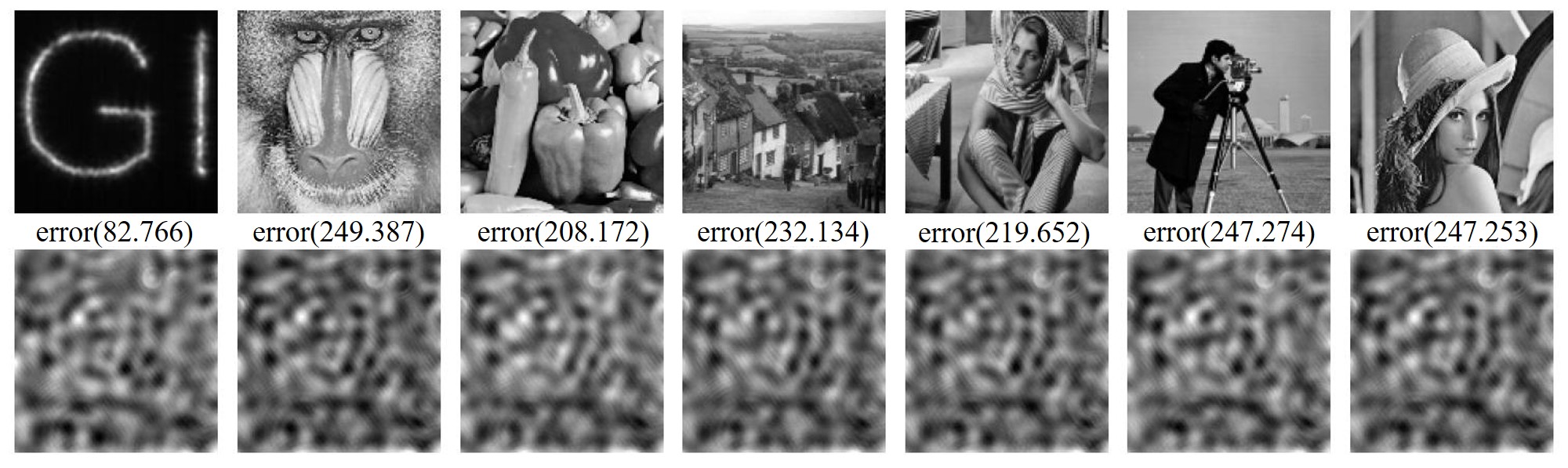}
  \caption{Mismatch recovery results without noise. The first line is ground truth images, and the second line is the restored images}.
  \label{fig:mismatch}
\end{figure}

We collect $M=2500$ speckle patterns ($N=128*128$) from multimode fiber at an offset of 25 as unknown measurement matrix $A_u$, speckle patterns collected from multimode fiber at an offset of 0 as measurement matrix $A$ for pre-measurement. We use GPSR algorithm as compressed sensing recovery. 7 images $x'$ are selected for the experiments. And Experiments (Exps) are implemented with RTX 2080 Ti (11GB), Pytorch. During the experiments, we find that some of the experimental results are related to the precision of the device used, which we will discuss in the next section.
~\\\\
Firstly, we present the mismatch recovery results, which use the measurement values $y'$ obtained by unknown measurement matrix $A_u$ and the pre-measurement matrix $A$ as mismatch pair $(y', A)$ input into the GPSR algorithm. In order to evaluate the rationality of the matrix constructed by the algorithm, the following error is defined
$$error = E||y'-A_{recv}x'||$$
We can see mismatch pair cann't recover the original images and its error is also large by \hyperref[fig:mismatch]{Fig.\ref{fig:mismatch}}, although the measurement is no noise ($\epsilon=0$).

\begin{figure}[htbp]
  \centering
  \includegraphics[scale = 0.335]{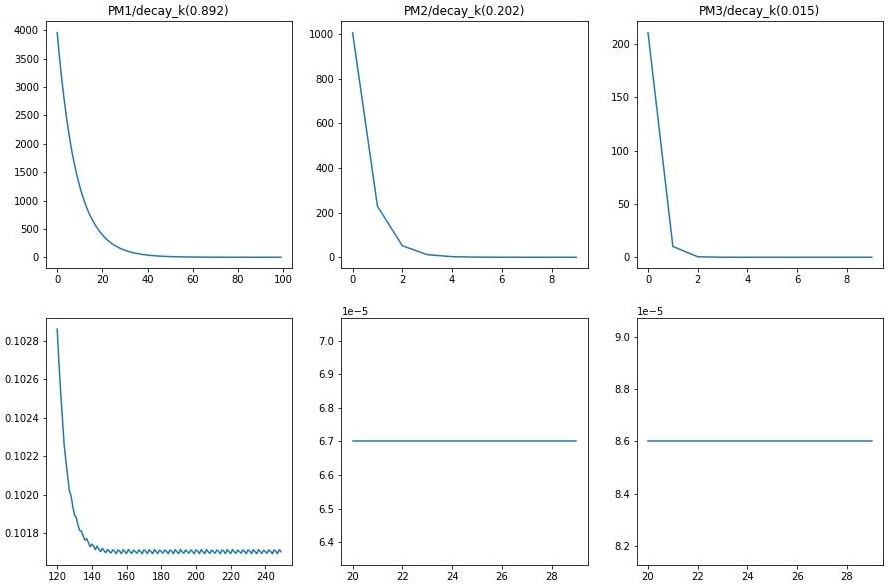}
\caption{Error curves of Algorithm.1. The horizontal axis is number of iteration and the vertical axis is error value. Same column shows the error curve of the same $PM_{image}$ in different iteration intervals.}.
  \label{fig:algo1-curve}
\end{figure}

\begin{figure}[H]
  \centering
  \setlength{\abovecaptionskip}{-0.05cm}
  \includegraphics[scale = 0.335]{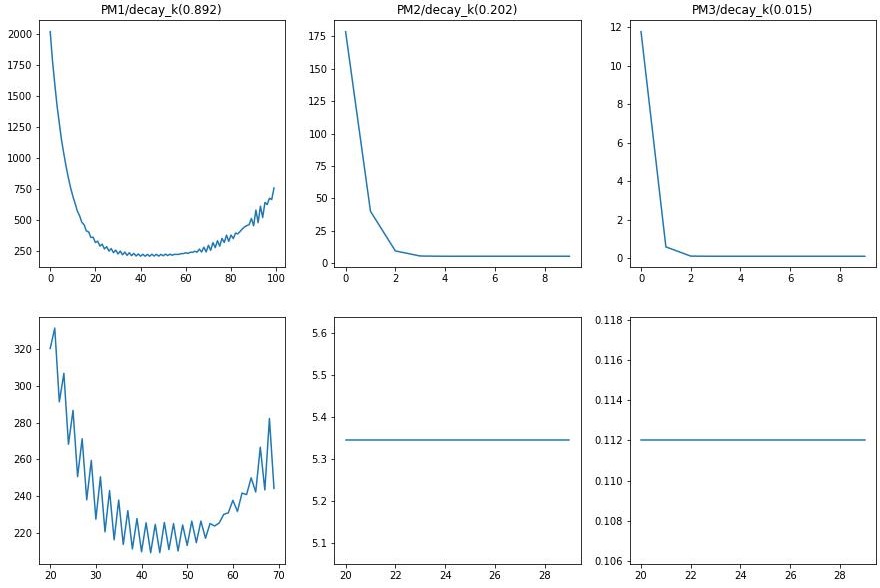}
\caption{Error curves of Algorithm.2. The horizontal axis is number of iteration and the vertical axis is error value. Same column shows the error curve of the same $PM_{image}$ in different iteration intervals.}.
  \label{fig:algo2-curve}
\end{figure}

\begin{figure}[H]
  \centering
  \setlength{\abovecaptionskip}{-0.05cm}
  \includegraphics[scale = 0.6]{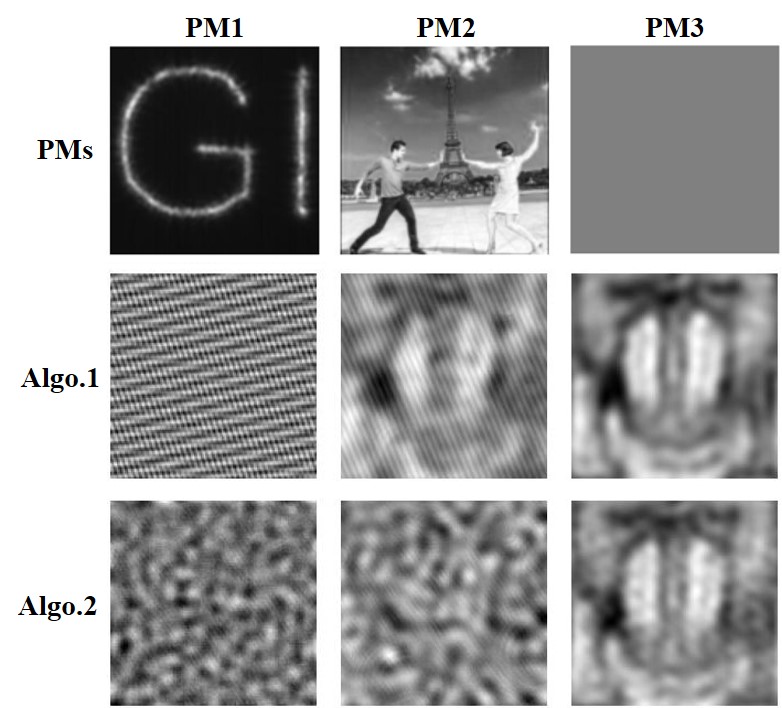}
  \caption{Restored Baboon image using $A_{recv}$ constructed by Algorithm.1 and Algorithm.2 under three different $PM_{image}$. PM1 is GI image. PM2 is a image of \textit{lsun/tower} dataset. PM3 is gray image of 0.5.}.
  \label{fig:recv-res}
\end{figure}

\begin{figure}[H]
  \centering
  \setlength{\abovecaptionskip}{-0.05cm}
  \includegraphics[scale = 0.23]{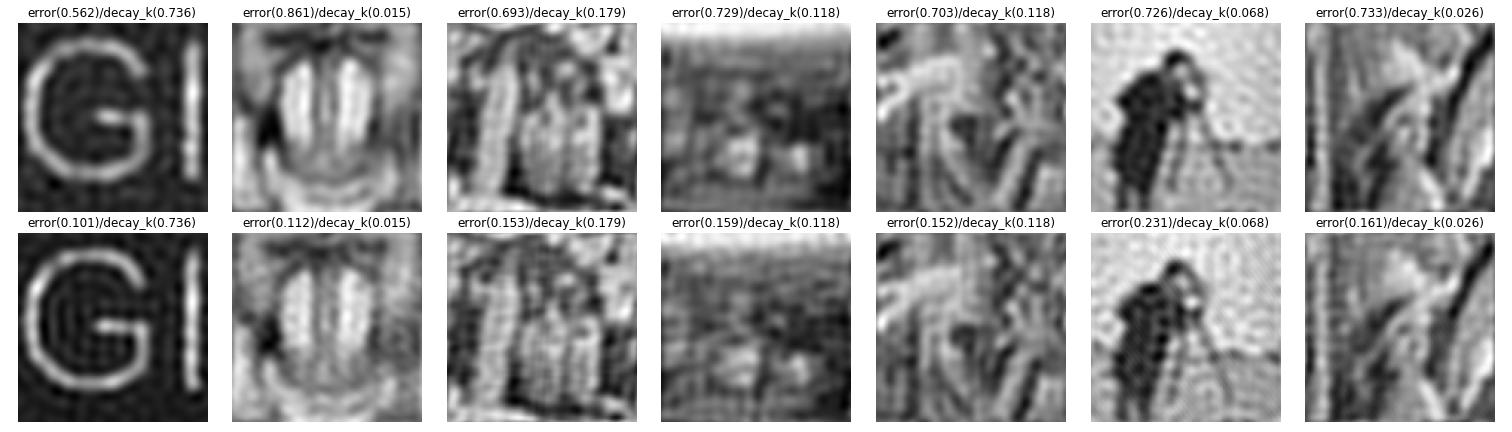}
\caption{Restored seven images using $A_{recv}$ constructed by Algorithm.1 (first line and it's error need multiple 1e-4) and Algorithm.2 (second line) under PM3.}.
  \label{fig:recv-res-row1}
\end{figure}

\begin{figure}[H]
  \vspace{-0.5pt}
  \centering
  \setlength{\abovecaptionskip}{-0.05cm}
  \includegraphics[scale = 0.44]{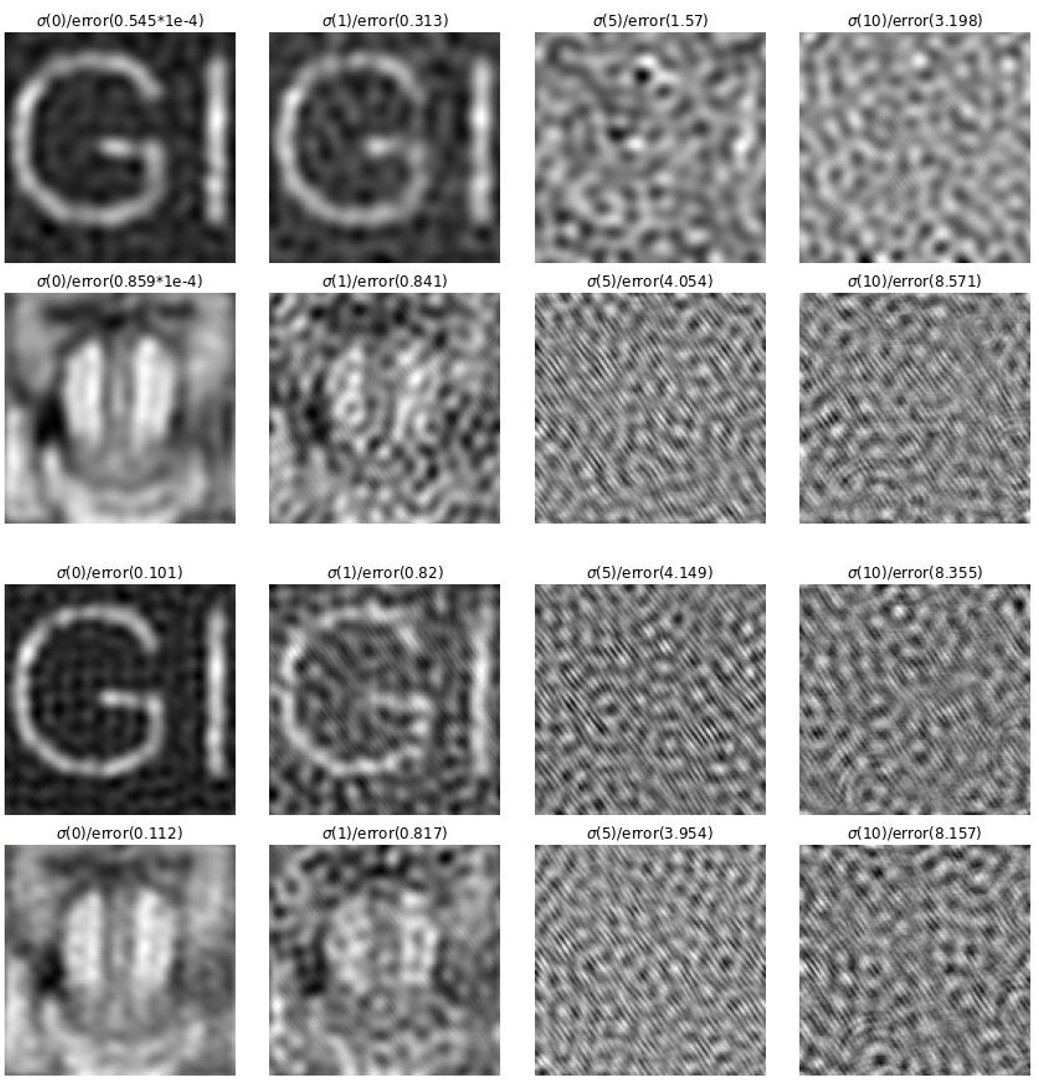}
\caption{Restored GI and Baboon images using $A_{recv}$ constructed by Algorithm.1 (1-2 lines) and Algorithm.2 (3-4 lines) under PM3. Same column has the same noise level. From left to right, different column has different noise level, which $\sigma$ is 0,1,5,10 in sequence.}.
  \label{fig:noise}
\end{figure}

\begin{figure}[H]
  \centering
  \setlength{\abovecaptionskip}{-0.05cm}
  \includegraphics[scale = 0.26]{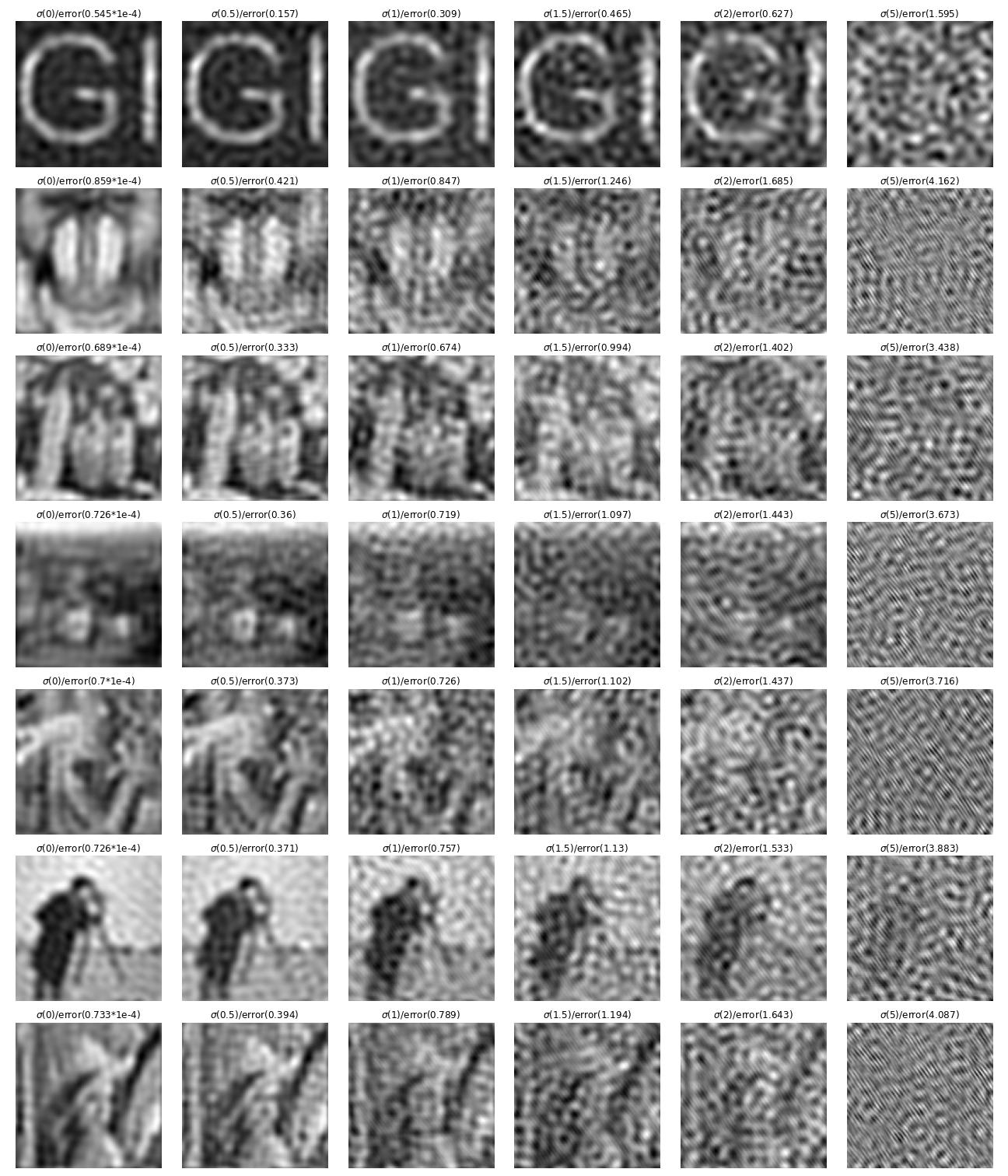}
\caption{Restored seven images using $A_{recv}$ constructed by Algorithm.1 under PM3. Same column has the same noise level. From left to right, different column has different noise level, which $\sigma$ is 0,0.5,1,1.5,2,5 in sequence.}.
  \label{fig:recv-res-algo1}
\end{figure}

\begin{figure}[H]
  \centering
  \setlength{\abovecaptionskip}{-0.05cm}
  \includegraphics[scale = 0.26]{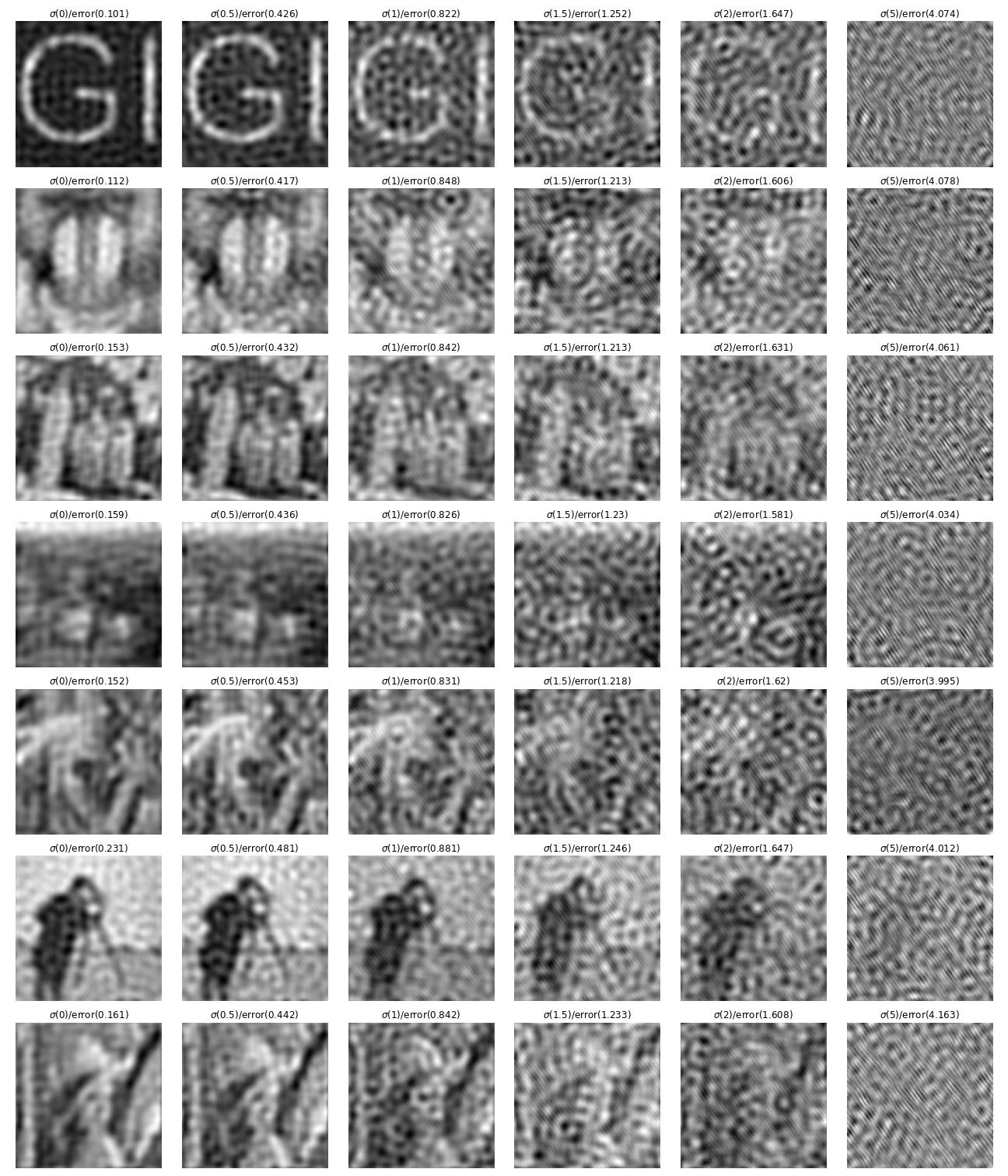}
\caption{Restored seven images using $A_{recv}$ constructed by Algorithm.2 under PM3. Same column has the same noise level. From left to right, different column has different noise level, which $\sigma$ is 0,0.5,1,1.5,2,5 in sequence.}.
  \label{fig:recv-res-algo2}
\end{figure}

\textbf{Exp0}: We attempt to use 3 different $PM_{image}$ for \hyperref[algo:1]{Algorithm.\ref{algo:1}} and \hyperref[algo:2]{Algorithm.\ref{algo:2}} as shown in \hyperref[fig:recv-res]{Fig.\ref{fig:recv-res}}, and plot error curves throughout the iteration process. This experiment is still no noise. In algorithm.1 as \hyperref[fig:algo1-curve]{Fig.\ref{fig:algo1-curve}}, using PM2 and PM3, the error converges to a very low value 1e-5 after 20 iterations, while using PM1 requires 140 iterations to converge to 0.1. And $decay\_k$ is $k_{\epsilon}$ as \hyperref[eq:itercond]{Eq.\ref{eq:itercond}}. We can see the closer it approaches to zero, the better convergence. That's exactly what happened in the experiment, which $decay\_k$ order is $PM1>PM2>PM3\rightarrow 0$. In algorithm.2 as \hyperref[fig:algo2-curve]{Fig.\ref{fig:algo2-curve}}, we can get the same results. The difference is that it does not converge when using PM1, showing a decrease followed by an increase. And the convergence error using PM2 and PM3 is higher than algorithm.1.
\newpage
From the final recovery results of Baboon image in \hyperref[fig:recv-res]{Fig.\ref{fig:recv-res}}, PM3 has the best performance. Notice in algo.2, the error of using PM2 ultimately converges to 5.35, while using PM3 converges to 0.112. Therefore, the former cann't recover, while the latter can.
\textbf{Exp1}: We attempt to use \hyperref[algo:1]{Algorithm.\ref{algo:1}} and \hyperref[algo:2]{Algorithm.\ref{algo:2}} to recover seven images using PM3 as \hyperref[fig:recv-res-row1]{Fig.\ref{fig:recv-res-row1}}. This experiment is still no noise. Original images have been successfully recovered.
\newpage
\textbf{Exp2}: We attempt to add following gaussian noise $\epsilon$ for each $speckle\_measure$. In \hyperref[fig:noise]{Fig.\ref{fig:noise}}, we can see that images still can be recovered when standard deviation $\sigma$ of noise is 1 and error is about 0.8. But images cann't be recovered when it is 5 and error is about 4. It indicates that constructed $A_{recv}$ does not have strong robustness to noise. More detailed noise level division and experimental results of all seven images are shown as \hyperref[fig:recv-res-algo1]{Fig.\ref{fig:recv-res-algo1}} and \hyperref[fig:recv-res-algo2]{Fig.\ref{fig:recv-res-algo2}}, which as additional supplementary in \hyperref[appe:D]{Appendix.\ref{appe:D}}.
$$\epsilon = \sigma * N(0,1)$$

\begin{figure}[H]
  \centering
  \setlength{\abovecaptionskip}{-0.05cm}
  \includegraphics[scale = 0.225]{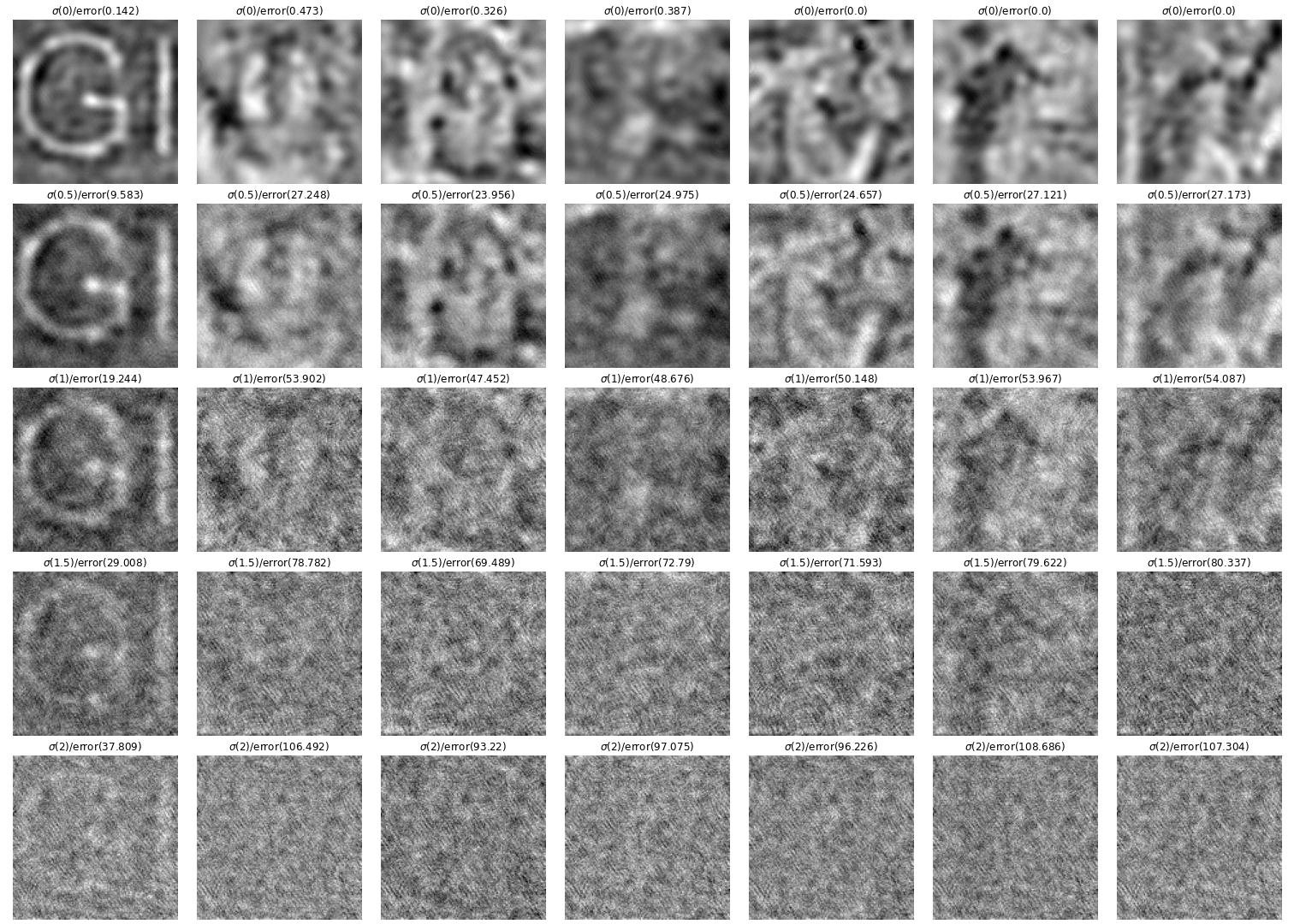}
\caption{Restored images using $A_{recv}$ constructed by Algorithm.4. The last three images are in $\mathcal{S}$ space, which error is zero without noise. The first four images are outside $\mathcal{S}$ space, which error is about 0.4 without noise. Same row has the same noise level. From top to bottom, different row has different noise level, which $\sigma$ is 0,0.5,1,1.5,2 in sequence.}.
  \label{fig:calibrate-res}
\end{figure}

\textbf{Exp3}: We attempt to calibrate unknown measurement matrix by \hyperref[algo:calidate-mspace]{Algorithm.\ref{algo:calidate-mspace}}. Construct orthogonal matrix $Q\in R^{N\times M}$ by qr-decomposition of the transposition of pre-measurement matrix $A^T$. So it's column vectors are base vectors. In order to make some images in the space $\mathcal{S}$ composed of the column vectors of the orthogonal matrix $Q$, we need to replace some columns of $A$ with the image vectors. We choose the last three images are in $\mathcal{S}$ space, while the first four images are outside $\mathcal{S}$ space. In \hyperref[fig:calibrate-res]{Fig.\ref{fig:calibrate-res}}, we can see that calibrated $A_{recv}$ can barely restore the original image. And it is also not robust to noise.

\subsection{The Impact of Device Precision}
\begin{figure}[H]
  \centering
  \setlength{\abovecaptionskip}{-0.05cm}
  \includegraphics[scale = 0.42]{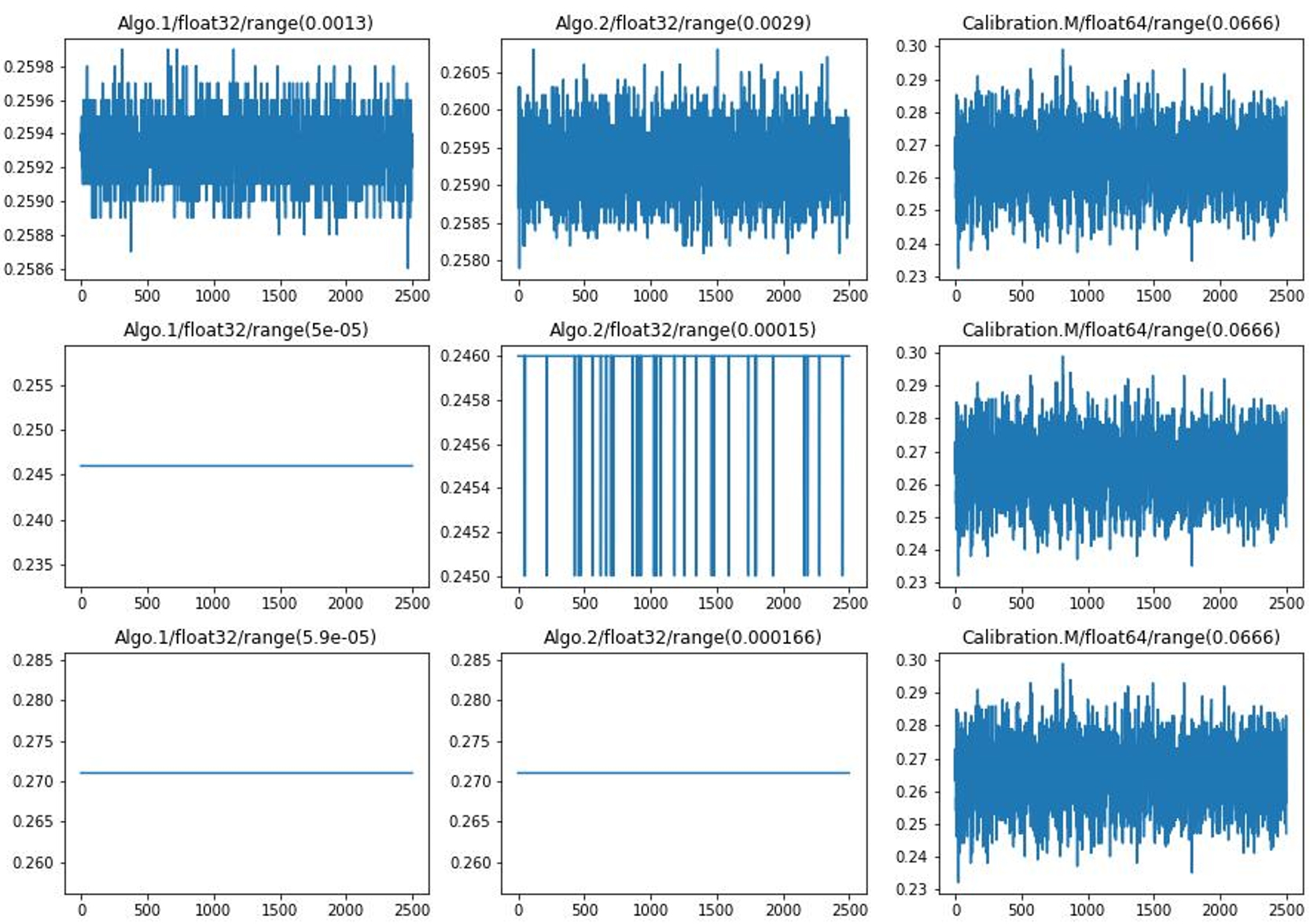}
\caption{$\vec{\lambda}$ components curves of three algorithms in three devices. From left to right, the columns represent Algorithm.1, Algorithm.2 and Algorithm.3. From top to bottom, the rows represent RTX2080 Ti, RTXA400, RTX3090. Algorithm.1 and Algorithm.2 calculated by float32, algorithm.3 calculated by float64. Range variable represents the range of components, which is maximum component minus 
minimum component.}
  \label{fig:mt-curve}
\end{figure}

To understand the impact of device precision, we need to get the core step of \hyperref[algo:1]{Algorithm.\ref{algo:1}}, \hyperref[algo:2]{Algorithm.\ref{algo:2}} and \hyperref[algo:calidate-mspace]{Algorithm.\ref{algo:calidate-mspace}}. It is that final constructed $A_{recv}$ is the result of $A_{recv}^{e_y^k}$ accumulation
\begin{gather*}
A_{recv} = \sum_{k} A_{recv}^{e_y^k} \\
y' \approx y_{recv} = A_{recv}x' = \sum_{k} \frac{y_0^T\Sigma Ax'}{y_0^T\Sigma y_0}\;e_y^k = \sum_{k} \lambda(y_0,\Sigma,A,x') \;e_y^k \\
\lambda(y_0,\Sigma,A,x') \in R
\end{gather*}
Furthermore, can we extract $\lambda(y_0,\Sigma,A,x')$ out of sum? The answer is that we can extract in Algorithm.1 and Algorithm.2, but we cann't extract in Algorithm.3. In the former two algorithms, parameters $(y_0,\Sigma,A,x')$ in $\lambda$ are all not related to iteration variable $k$. In the latter algorithm, parameters $y_0$ in $\lambda$ is related to iteration variable $k$, which $e_y^k=Y_u(:,k)$ and $y_0=Y(:,k)$. Now we consider the former two algorithms, parameters $(y_0,\Sigma,A)$ in $\lambda$ not related to parameter $x'$. If we can only consider $x'$ as a variable and the rest as constants, we have
\begin{gather*}
y' \approx A_{recv}x'=\lambda(y_0,\Sigma,A,x')\sum_{k}e_y^k=\lambda(x')\sum_{k}e_y^k 
\end{gather*}
This means the feasible solution of pair $(y',A_{recv})$ input into  compressed sensing algorithm can be any image $x$ because of multiplier property of mismatch equation mentioned above as \hyperref[eq:m]{Eq.\ref{eq:m}}.
\begin{gather}
\label{eq:3-1}
A_{recv}x=\lambda(x)\sum_{k}e_y^k=\frac{\lambda(x)}{\lambda(x')}\lambda(x')\sum_{k}e_y^k \approx \frac{\lambda(x)}{\lambda(x')}y' 
\end{gather}
Now if we do the former two algorithms in the device with high precision, the coefficient $\lambda(x)/\lambda(x')$ is the constant, which does not affect the feasible solution is any image $x$. And most compression sensing algorithms choose the sparsest solution as the optimal solution, which is the image filled with constant grayscale. But if we do the former two algorithms in the device with low precision, constructed measurement matrix $A_{recv}$ does not strictly meet the multiplier property. This is to say we have
\begin{subequations}
\begin{gather}
\label{eq:3-2.a}
y' \approx A_{recv}x' \tag{3-2.a} \\
\label{eq:3-2.b}
A_{recv}x = \vec{\lambda} \star (A_{recv}x') \approx \vec{\lambda} \star y' \tag{3-2.b}
\end{gather}
\end{subequations}
$\star$ represents point-wise multiplication of vectors. $\vec{\lambda}$ is a vector with the same dimension of $y'$ and it's components have fluctuation, not constant. And \hyperref[eq:3-2.a]{Eq.\ref{eq:3-2.a}} is still hold. So in this case, pair $(y',A_{recv})$ input into compressed sensing algorithm can get optimal solution $x'$ without any image $x$ confusion. And components of $\vec{\lambda}$ have the larger fluctuation, the lower any image confusion and the more exact optimal solution $x'$.
~\\\\
As for latter Algorithm.3, we cann't extract $\lambda(y_0,\Sigma,A,x')$ out of sum. Obviously, it doesn't have  \hyperref[eq:3-1]{Eq.\ref{eq:3-1}} hold but rather have \hyperref[eq:3-2.b]{Eq.\ref{eq:3-2.b}} hold. So Algorithm.3 is not impacted by device precision.
~\\\\
To verify the appeal theory, we conduct the same experiment on three different devices, which are RTX2080 Ti, RTXA400, RTX3090. The same experiment is that \hyperref[algo:1]{Algorithm.\ref{algo:1}} and \hyperref[algo:2]{Algorithm.\ref{algo:2}} using GI image as unknown images $x'$ and other parameters $(PM3, A, A_u)$ remained consistent with previous experiments to construct $A_{recv}$, \hyperref[algo:calidate-mspace]{Algorithm.\ref{algo:calidate-mspace}} directly uses same parameters $(A, A_u)$ to construct $A_{recv}$. Then select Baboon image as other image $x$ and put pair $(A_{recv},x',x)$ into \hyperref[eq:3-2.b]{Eq.\ref{eq:3-2.b}} calculating $\vec{\lambda}$, which components curves as shown in \hyperref[fig:mt-curve]{Fig.\ref{fig:mt-curve}}. From it, we can know
the fluctuation and range of $\vec{\lambda}$ components in Algorithm.1 and Algorithm.2, which strength order of device is RTX2080 Ti\textgreater RTXA4000\textgreater RTX3090. But in Algorithm.3, there are the same large fluctuation and range on all devices.

\begin{figure}[H]
  \centering
  \setlength{\abovecaptionskip}{-0.05cm}
  \includegraphics[scale = 0.7]{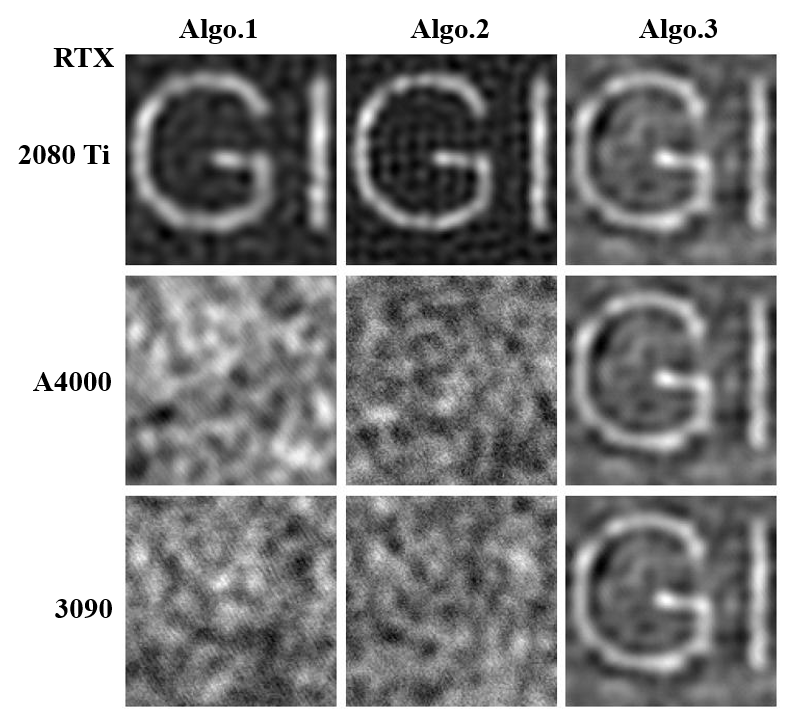}
\caption{Restored GI image using $A_{recv}$ constructed by three algorithms under PM3 in three devices.}
  \label{fig:mt-res}
\end{figure}

And the restored unknown images $x'$ as shown in \hyperref[fig:mt-res]{Fig.\ref{fig:mt-res}}, we can get the consistent conclusion. When Algorithm.1 and Algorithm.2 are on RTX2080 Ti device, which have large fluctuation and range of $\vec{\lambda}$ components, the unknown images can be restored. But when they are on RTXA4000 and RTX3090 devices, which have very small fluctuation and range of $\vec{\lambda}$ components and even constant, the unknown images cann't be restored. And Algorithm.3 on all devices, the unknown images can be restored.
~\\\\
Based on appeal analysis and experiments, we can summarize that Algorithm.1 and Algorithm.2 should use float32 for calculations on medium precision devices, while Algorithm.3 should use float64 for calculations on higher precision devices as much as possible.

\section{Conclusions}
We propose two types of method to transform the mismatch problem into matched, which are matched solution of unknown measurement matrix and calibration of unknown measurement matrix. And based on mismatch equation, three algorithms are provided to construct 
matrix $A_{recv}$. Both theoretical analysis and experimental results express that constructed matrix can replace unknown measurement matrix to restore the original images. The core idea of algorithm in first type is using one or more known measurement to gain on the measurement value of unknown measurement matrix which is only applicable to the measurement value of current image, while algorithm in second type is using M base images of a specific space to calibrate matrix which is applicable to the measurement value of any image in that space. Experiments have shown that at low noise levels, the constructed matrix $A_{recv}$ can restore the original images. Additionally, we explain the impact of device calculation precision on the three algorithms by detailed analysis of the multiplier property of mismatch equation.
\newpage

\bibliographystyle{unsrt} 
\bibliography{cites}

\newpage
\appendices
\section{Proof of Mismatch Equation}
\label{appe:A}

Considering without noise, known and unknown measurements are as following\begin{subequations}
\label{eq:af}
\begin{align}
\label{eq:a1}
y_0&=A x \tag{a-1} \\
\label{eq:a2}
y&=A_u x \tag{a-2}
\end{align} 
\end{subequations}We can easily verify $A_{recv}^{(y_0,y)}$ is a general solution.

\begin{equation}
A_{recv}^{(y_0,y)} = \frac{1}{y_0^T\Sigma y_0}yy_0^T\Sigma A \nonumber
\end{equation}

\begin{equation}
A_{recv}^{(y_0,y)}x = \frac{1}{y_0^T\Sigma y_0}yy_0^T\Sigma A x = \frac{y_0^T\Sigma A x}{y_0^T\Sigma y_0} y = \frac{y_0^T\Sigma y_0}{y_0^T\Sigma y_0} y = y \nonumber
\end{equation}

Now we are trying to directly solve a special solution from \hyperref[eq:af]{Eq.a}. Assuming the existence of matrix $C\in R^{N\times M}$, making $CA$   invertible, then \hyperref[eq:a1]{Eq.\ref{eq:a1}} multiply both sides by $C$, we have
\begin{gather*}
Cy_0=CA x \\
(CA)^{-1}Cy_0=x 
\end{gather*}
Then, substitute the results into \hyperref[eq:a2]{Eq.\ref{eq:a2}}, we have
\begin{equation}
\label{eq:asenc}
y=A_ux=A_u(CA)^{-1}(Cy_0) \tag{a-3}   
\end{equation}

Assuming the existence of matrix $D\in R^{1\times N}$, making $Cy_0D$ invertible, then \hyperref[eq:asenc]{Eq.\ref{eq:asenc}} multiply both sides by $D$, we have
\begin{gather*}
yD=A_u(CA)^{-1}(Cy_0D) \\
yD(Cy_0D)^{-1}(CA)=A_u
\end{gather*}
Since we solve the special solution, we can assume that invertible condition is orthogonal. So let's summarize the current assumptions

\begin{description}
    \item[1)] $Cy_0D$ is orthogonal
\end{description}

\begin{equation}
\begin{aligned} 
A_u&=yD(Cy_0D)^{-1}(CA)  \\
&=yD(Cy_0D)^{T}(CA) \\
&=yDD^Ty_0^TC^TCA  \\
&=||D|| yy_0^TC^TCA \nonumber
\end{aligned}
\end{equation}

Now we need to solve $\Omega=C^TC$, we need to use the second assumption
\newpage
\begin{description}
    \item[2)] $CA$ is orthogonal
\end{description}

Because $A\in R^{M\times N}$ and $M\ll N$ is row full rank, $AA^T$ is invertible, we can know $\Omega=(AA^T)^{-1}$ from \hyperref[lam:1]{Lemma.\ref{lam:1}} . Now we have
\begin{gather*}
A_u = ||D|| yy_0^T(AA^T)^{-1}A
\end{gather*}
Now let's rephrase $A_u$ back to \hyperref[eq:a2]{Eq.\ref{eq:a2}}. we can get coefficients $||D||$
\begin{gather*}
y=A_ux = ||D|| yy_0^T(AA^T)^{-1}Ax=||D||yy_0^T(AA^T)^{-1}y_0 \\
||D|| = \frac{1}{y_0^T(AA^T)^{-1}y_0}
\end{gather*}

So we get the special solution
\begin{gather*}
A_u =\frac{1}{y_0^T(AA^T)^{-1}y_0} yy_0^T(AA^T)^{-1}A
\end{gather*}

Since $\Sigma=(AA^T)^{-1}$ is a matrix of $R^{M\times M}$, replacing it with another matrix does not change the result $y=A_ux$. Therefore, we can consider the general solution to be \hyperref[eq:Arecv]{Eq.\ref{eq:Arecv}} as \textbf{Mismatch Equation}.
~\\\\
\textbf{Lemma 1.} \textit{If $CA$ is orthogonal and $AA^T$ is invertible, we have}
\begin{equation}
\label{lam:1}
C^TC=(AA^T)^{-1} \tag{1}
\end{equation}
\textit{Proof.} By the definition of $A\in R^{M\times N}$ and $C\in R^{N\times M}$, we have $CA\in R^{N\times N}$ is square matrix, we have
\begin{gather*}
\begin{aligned}
AA^T = AEA^T = A(CA)^T&CAA^T = AA^TC^TCAA^T \\
\Longrightarrow C^TC = & \;(AA^T)^{-1}
\end{aligned}
\end{gather*}

\section{Convergence Proof of Iterative Algorithm}
\label{appe:B}

Considering with noise, we can proof the convergence of \hyperref[algo:1]{Algorithm.\ref{algo:1}}. Firstly, it's mathematical description as following
\begin{subequations}
\begin{align}
&A_{recv}^0=0,  \;\; e_y^0=y', \;\; \epsilon^0=0 \notag \\
\label{eq:b1}
& A_{recv}^{k+1} = A_{recv}^k + A_{recv}^{e_y^k} \tag{b-1}\\
\label{eq:b2}
& e_y^{k}=y'-A_{recv}^{k}x'-\epsilon^{k} \tag{b-2}
\end{align}
\end{subequations}
The constructed matrix for the final output of the iteration is $A_{recv} = \lim\limits_{k\to+\infty} A_{recv}^{k}$. Let $y_{recv}^{k} = A_{recv}^{k}x'$ and $\lambda^{k} = y'-y_{recv}^{k}$. So our goal is $\lambda^{k}$ can convergent. The most ideal situation is converge to zero.
\begin{gather*}
\begin{aligned}
\lim\limits_{k\to+\infty}{(y'-A_{recv}^{k}x')}&=0    \\
\lim\limits_{k\to+\infty}{(y'-y_{recv}^{k})}&=0 \\
\lim\limits_{k\to+\infty}{\lambda^{k}}&=0 \\
\end{aligned}
\end{gather*}

\begin{figure}[H]
  \centering
  \setlength{\abovecaptionskip}{-0.05cm}
  \includegraphics[scale = 0.65]{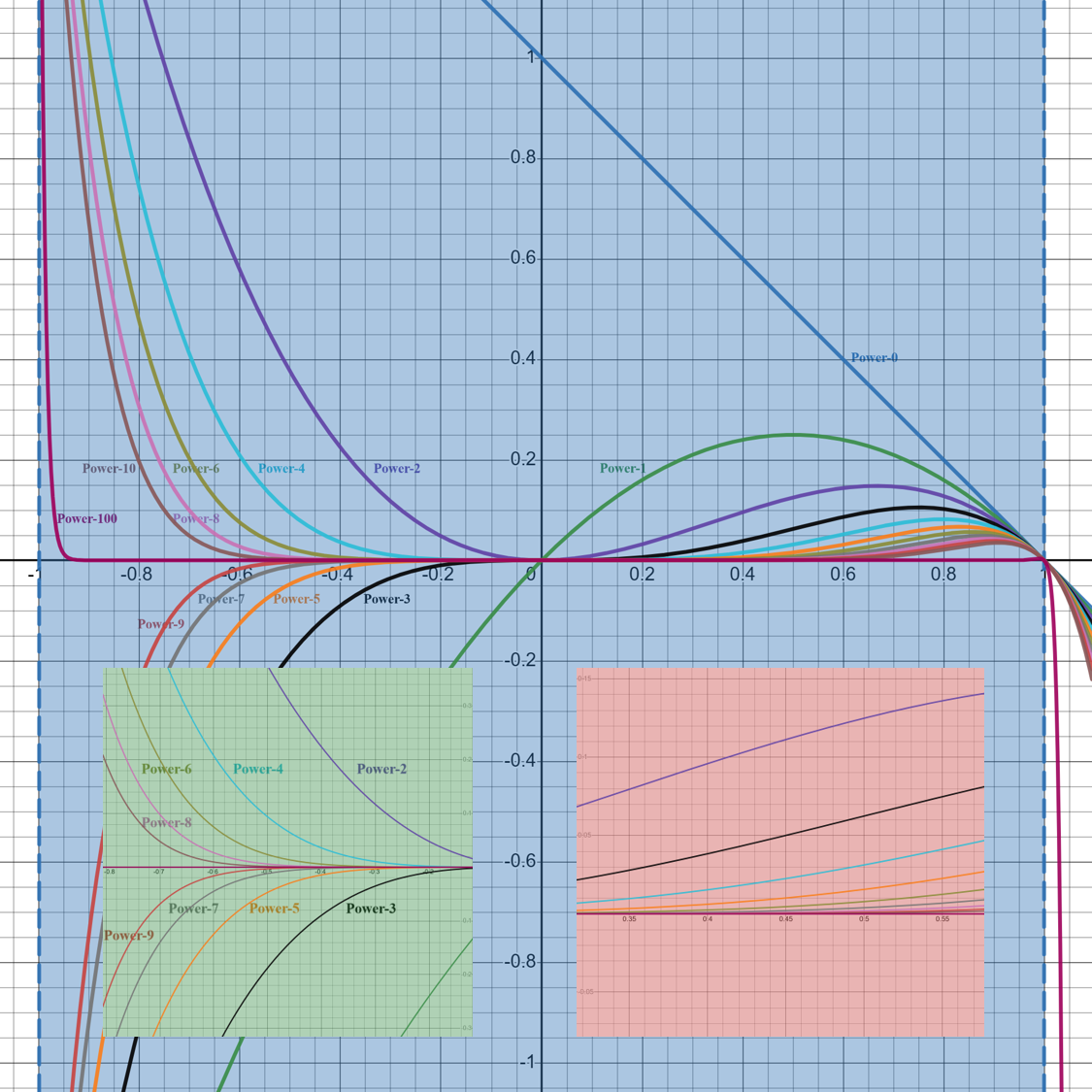}
\caption{Curve family of $\{(1-x)x^{i}\;\vert \;Abs(x)\textless 1\}_{i}$. Left green sub-figure is a scaled figure of parent figure centered at x point of -0.45. Right red sub-figure is a scaled figure of parent figure centered at x point of 0.45.}.
  \label{fig:powers}
\end{figure}

For the convenience of subsequent derivation, we will perform some deformation processing on the pre-measure
\begin{gather*}
\begin{aligned}
y_0=A*PM_{image} = &\;Ax'+A(PM_{image} - x') =Ax'+\epsilon \\ 
\Longrightarrow \epsilon = &\;A(PM_{image} - x')
\end{aligned}
\end{gather*}
Secondly, let's derive a recurrence formula for $\lambda^{k}$. We can do following simplification by \hyperref[eq:b1]{Eq.\ref{eq:b1}}
\begin{gather*}
\begin{aligned}
y_{recv}^{k+1} = A_{recv}^{k+1}x' &= A_{recv}^kx' + A_{recv}^{e_y^k}x' \\
&=A_{recv}^kx' + \frac{1}{y_0^T\Sigma y_0} e_y^k y_0^T\Sigma Ax' \\
&=A_{recv}^kx' + \frac{1}{y_0^T\Sigma y_0} e_y^k y_0^T\Sigma (y_0-\epsilon) \\
&=A_{recv}^kx' + \frac{1}{y_0^T\Sigma y_0} e_y^k y_0^T\Sigma y_0 -\frac{1}{y_0^T\Sigma y_0} e_y^k y_0^T\Sigma \epsilon \\
&=A_{recv}^kx' + e_y^k -\frac{y_0^T\Sigma \epsilon}{y_0^T\Sigma y_0} e_y^k \\
&=A_{recv}^kx' + (1-k_{\epsilon})e_y^k  \\
&=y_{recv}^{k} + (1-k_{\epsilon})e_y^k  \\
\Longrightarrow k_{\epsilon}&=\frac{y_0^T\Sigma \epsilon}{y_0^T\Sigma y_0}
\end{aligned}
\end{gather*}
Using \hyperref[eq:b2]{Eq.\ref{eq:b2}} to eliminate $e_{y}^k$, we have
\begin{subequations}
\begin{align*}
e_y^{k}&=y'-A_{recv}^{k}x'-\epsilon^{k} \\
&=y'-y_{recv}^{k}-\epsilon^{k} \\ 
\Longrightarrow y_{recv}^{k+1} &= y_{recv}^k+(1-k_{\epsilon})e_y^k \\
&= y_{recv}^k+(1-k_{\epsilon})(y'-y_{recv}^k-\epsilon^{k}) \\
&=y_{recv}^k+y'-y_{recv}^k-\epsilon^{k}-k_{\epsilon} * (y'-y_{recv}^k-\epsilon^{k}) \\
y'-y_{recv}^{k+1} &=\epsilon^{k}+k_{\epsilon} * (y'-y_{recv}^k-\epsilon^{k}) \\
y'-y_{recv}^{k+1} &=k_{\epsilon}(y'-y_{recv}^k)+(1-k_{\epsilon})\epsilon^{k} \\
\label{eq:recformula}
\lambda^{k+1}&=k_{\epsilon}\lambda^{k}+(1-k_{\epsilon})\epsilon^{k} \tag{b-3}
\end{align*}
\end{subequations}
From \hyperref[eq:recformula]{Eq.\ref{eq:recformula}}, we can obtain the general term expression of the sequence $\lambda^k$
\begin{equation}
\label{eq:finequ}
\lambda^k = (k_{\epsilon})^k\lambda^0 + (1-k_{\epsilon})\sum_{i=0}^{k-1}(k_{\epsilon})^{i}\epsilon^{k-i} \tag{b-4}
\end{equation}
In order to converge, we must select appropriate special solutions $\Sigma$ and pre-measure $y_0$, making
\begin{equation}
\label{eq:itercond}
||k_{\epsilon}||=||\frac{y_0^T\Sigma \epsilon}{y_0^T\Sigma y_0}||=\frac{||y_0^T\Sigma A(PM_{image}-x')||}{||y_0^T\Sigma y_0||}=||1-\frac{y_0^T\Sigma Ax'}{y_0^T\Sigma y_0}||<1 \tag{b-5}
\end{equation}
Assuming that the measurement noise each time is \textit{i.i.d.}, we have $\epsilon^{k}\sim N(\mu,\sigma^2)$
$$
\lim\limits_{k\to+\infty}{\lambda^{k}}\sim N(\mu,\frac{1}{k_{\epsilon}+1}\sigma^2)
$$
Specifically, when there is no noise, the limit tends to zero. When noise is added, the algorithm can converge at the same noise level. And From \hyperref[eq:finequ]{Eq.\ref{eq:finequ}}, we can know the best choise of $k_{\epsilon}$ to minimize the impact of noise is in the neighbourhood of 0. This is to say
\begin{gather*}
When\;\; k_{\epsilon} \in U(0,\delta_1) \;\; and \;\; i>0 \Longrightarrow (1-k_{\epsilon})(k_{\epsilon})^i \in U(0,\delta_2) \;\; and \;\; \delta_2 \rightarrow 0
\end{gather*}
As shown in \hyperref[fig:powers]{Fig.\ref{fig:powers}}, we can know that the larger $i$, the wider neighbourhood radius $\delta_1$. But when $i=0$, noise coefficient degenerates into $1-k_{\epsilon}$, which in $U(1,\delta_2)$ making it impossible to eliminate the last noise $\epsilon^{k}$. 

\section{Proof of Calibration Equation}
\label{appe:C}

Assuming unknown image $x'$ can be linearly represented by orthogonal base images $\{\bf{x_i}\}_{i=1}^{N}$ in space
$$x'=\sum_{i}^{N} b_i\bf{x_i}$$
Perform Pre-Measure and Unknown-Measure on all base images
\begin{gather*}
y_{i}^0=A\mathbf{x_i}\\
y_{i}=A_u\mathbf{x_i}+\epsilon_{i}     
\end{gather*}
We can proof that $A_{recv}$ as following equation is an exact solution for unknown measurement matrix $A_u$
$$
A_{recv}=\sum_{j}^{N} a_j A_{recv}^{(y_{j}^0,y_j)}
$$
Calculate the expected measurement value for image $x'$ under above $A_{recv}$ 
$$y_{recv} = A_{recv} x' = \sum_{i}^{N} b_iA_{recv} \mathbf{x_i} = \sum_{i}^{N}b_i\sum_{j}^{N}a_jA_{recv}^{(y_{j}^0,y_j)}\mathbf{x_i}$$
Further simplification based on the following multiple properties
\begin{gather*}
\begin{aligned}
A_{recv}^{(y_{j}^0,y_j)}\mathbf{x_i} &= \frac{1}{(y_j^0)^T\Sigma  y_j^0} y_j(y_j^0)^T\Sigma A\mathbf{x_i}=k(i,j)y_j \\
\Longrightarrow k(i,j)&= \frac{(y_j^0)^T\Sigma A\mathbf{x_i}}{(y_j^0)^T\Sigma  y_j^0} = \frac{(y_j^0)^T\Sigma y_i^0}{(y_j^0)^T\Sigma  y_j^0}
\end{aligned}
\end{gather*}
$$
y_{recv} = A_{recv} x' = \sum_{i}^{N}b_i\sum_{j}^{N} a_jk(i,j)y_j = \sum_{j}^{N} (\sum_{i}^{N}b_i*a_j*k(i,j)) y_j
$$
On the other hand, we have unknown measurement
$$
y = A_ux'+\epsilon= \sum_{i}^{N} b_i A_u\mathbf{x_i}+\epsilon= \sum_{i}^{N} b_i (y_i-\epsilon_{i})+\epsilon= \sum_{j}^{N} b_j y_j -\sum_{j}^{N} b_j \epsilon_{j} + \epsilon
$$
In order to make the difference between $y$ and $y_{recv}$ is within the error range, we just need to take the appropriate $\{a_j\}_{j=1}^{N}$ and $\Sigma$  satisfy the following condition
\begin{equation}
\label{eq:ccon}
\forall j \;\;\sum_{i}b_i*a_j*k(i,j)=b_j \tag{c-1}    
\end{equation}
So the error of the measurement values is
\begin{equation}
\label{eq:y-error}
y_{recv}-y=
\begin{cases}
\epsilon^{'} = \sum_{j}^{N} b_j \epsilon_{j} + \epsilon & nosie\\
0& without \;\; nosie
\end{cases} \tag{c-2}
\end{equation}
The only possible situation to satisfy the condition \hyperref[eq:ccon]{Eq.\ref{eq:ccon}} is
$$
a_j*k(i,j)=
\begin{cases}
1& \text{i=j}\\
0& \text{i$\neq$j}
\end{cases}
$$
If assume $\forall i\neq j\;\; k(i,j) \neq 0$. Since $k(i,i) \equiv 1$, sequence $\{a_j\}_{j=1}^{N}$ must satisfy the following condition
$$
a_j=\begin{cases}
1& \text{i=j}\\
0& \text{i$\neq$j}
\end{cases}
$$
Obviously this condition cannot be established because of $a_j$ cannot be equal to two values at the same time. So there is one situation for the choice of $\{a_j\}_{j=1}^{N}$ and $\Sigma$
\begin{gather*}
\forall j\;\;a_j\equiv 1\\
k(i,j)=\frac{(y_j^0)^T\Sigma y_i^0}{(y_j^0)^T\Sigma  y_j^0}=
\begin{cases}
1& \text{i=j}\\
0& \text{i$\neq$j}
\end{cases} 
\end{gather*}
Equivalent to
\begin{gather*}
(y_j^0)^T\Sigma y_i^0=
\begin{cases}
c\neq0& \text{i=j}\\
0& \text{i$\neq$j}
\end{cases} 
\end{gather*}
Arrange the pre-measurement value $\{y_j^0\}_{j=1}^{N}$ of all base images by rows to form a matrix $Y\in R^{N \times M}$, we can obtain a more concise representation for the choice of special solution $\Sigma$ in mismatch equation, where $E$ is the identity matrix
$$Y\Sigma Y^{T}=E$$

\section{Additional Supplementary}
\label{appe:D}

\end{document}